\documentstyle[aps,preprint]{revtex}
\begin{document}
\draft
\title{Control of Current Reversal in Single and Multiparticle Inertia Ratchets}
\author{H. A. Larrondo$^1$, Fereydoon Family$^2$, C. M. Arizmendi$^{1,2}$}
\address{$^1$ Depto. de F\'{\i}sica, Facultad de\\
Ingenier\'{\i}a, Universidad Nacional de Mar del\\
Plata,\\
Av. J.B. Justo 4302, 7600 Mar del Plata,\\
Argentina\\
$^2$Department of Physics, Emory\\
University, Atlanta, GA 30322, USA}
\date{\today }
\maketitle

\begin{abstract}
We have studied the deterministic dynamics of underdamped single and
multiparticle ratchets associated with current reversal, as a function of
the amplitude of the external driving force. Two experimentally inspired
methods are used. In the first method the same initial condition is used for
each new value of the amplitude. In the second method the last position and
velocity is used as the new initial condition when the amplitude is changed.
The two methods are found to be complementary for control of current
reversal, because the first one elucidates the existence of different
attractors and gives information about their basins of attraction, while the
second method, although history dependent, shows the locking process. We
show that control of current reversals in deterministic inertia ratchets is
possible as a consequence of a locking process associated with different
mean velocity attractors. An unlocking effect is produced when a chaos to
order transition limits the control range.
\end{abstract}


\bigskip

\newpage

Stochastic models known as {\it thermal}{\sl \ }{\it ratchets} or {\it %
correlation} {\it ratchets} \cite{general}, in which a nonzero net drift
velocity may be obtained from fluctuations interacting with broken symmetry
structures \cite{2}, have recently received much attention. This interest is
due to the possible applications of these models for understanding molecular
motors \cite{3}, nanoscale friction \cite{4}, surface smoothening \cite{5},
coupled Josephson junctions \cite{6}, optical ratchets and directed motion
of laser cooled atoms \cite{7}, and mass separation and trapping schemes at
the microscale \cite{8}. The fluctuations that produce the net transport are
usually associated with noise, but they may arise also in absence of noise,
with additive forcing, in overdamped deterministic systems \cite{Barbi7},
overdamped quenched systems\cite{quenched} and in underdamped systems \cite
{9,10,11,quenched2}.

Inertial ratchets, even in the absence of noise have a very complex
dynamics, including chaotic motion \cite{9,10}. This deterministically
induced chaos mimics, to some extent, the role of noise\cite{13a}, changing,
on the other hand, some of the basic properties of thermal ratchets. For
example, inertial ratchets can exhibit multiple reversals in the current
direction \cite{9,10}. The direction depends on the amount of friction and
inertia, which makes it especially interesting for technological
applications such as microscopic particle separation \cite{8}.

Jung {\it et al} \cite{9} studied the case of an underdamped particle
periodically driven in an asymmetric potential without noise and found
multiple current reversals varying with the intensity of the external
perturbation. They characterized the motion by cumulants of the contracted,
time-dependent solution of the Liouville equation and distinguished regular
from chaotic transport.

Several attempts to find the mechanism causing these current inversions have
been made. Mateos \cite{10} analyzed the relation between the bifurcation
diagram and the current. He conjectured that the origin of the current
reversal is the bifurcation from a chaotic to a periodic regime. Close to
this bifurcation he observed trajectories revealing intermittent chaos and
anomalous deterministic diffusion. Barbi{\it \ et al}. \cite{11} related the
transport properties to phase locking dynamics. They interpreted the current
reversals in terms of different stability properties of the periodic
rotating orbits and reported cases where current reversals appear also in
the absence of a bifurcation from a chaotic to a periodic motion. Although
the origin of current inversions seems to be relatively clear, this has not
been enough to propose a method to control the multiple current reversals
which is important technologically.

The aim of this paper is to elucidate the relationship between bifurcation
diagrams, phase locked dynamics and transport phenomena for an underdamped
deterministic ratchet without noise in order to find a way to control the
current reversals.

Specifically we study the dimensionless equation of motion: 
\begin{equation}
\epsilon \ddot{x}+\gamma \dot{x}=\cos (x)+\mu \cos (2x)+\Gamma \sin (\omega
t).
\end{equation}
Here, $\epsilon $ is the mass of the particle, $\gamma $ is the damping
coefficient, $\Gamma $ and $\omega $ are respectively the amplitude and
frequency of an external oscillatory forcing. The asymmetric potential is
given by

\[
U(x)=-\sin (x)-\frac \mu 2\sin (2x). 
\]

Numerical solutions of Eq. (1) are obtained using a variable step
Runge-Kutta-Fehlberg method\cite{12}. We fixed $\epsilon =1.1009$, $\gamma
=0.1109$, $\mu =0.5$ and $\omega =0.67$ and studied the behavior of the
system as a function of $\Gamma $. These parameters were chosen because they
are in the same region as used by Barbi and Salerno\cite{11}.

Let us start analyzing the case of only one particle with a specific initial
condition. The values of $x(t)$ and $v(t)$ are sampled with a sampling
period $T_s=T/20$ where $T$ is the period of the driving force (i.e. $T=2\pi
/\omega $).

The mean velocity of the particle is defined as:

\[
<v>=\frac{x(n_{\max }T)-x(n_{tran}T)}{(n_{\max }-n_{tran})T}, 
\]

where the number of periods of the transitory was chosen as $n_{tran}$=400
and the average was done until $n_{\max }$=500 periods.

There are at least two different ways to do a real experiment to study the
effect of the variation of the strength of the external force $\Gamma$. One
way is to let the particle evolve starting from the same initial condition
when $\Gamma$ changes. Another possible way is to change $\Gamma$ in the
middle of the trajectory of the particle. Simulations were carried out with
both methods, that will be called method I and II respectively,
corresponding to the two possible experimental realizations.

Let us first present the results of method I. The normalized mean velocity $%
<v>/v_\omega $ , with $v_\omega =T_x/T,$ where $T_x$ is the spatial period
of the potential is shown in figure 1a for a particle with initial condition
($x_0$=0, $v_0$=1) as a function of the amplitude of the external
oscillatory forcing $\Gamma$. The simulation was done by resetting the
initial condition to the same values ($x_0$=0, $v_0$=1) when $\Gamma$ was
changed. The corresponding bifurcation diagram is plotted in figure 1b.
Figures 1c and 1d are enlargements for $\Gamma \in \left[ 1,1.05\right] $.
There are several important remarks concerning these figures: a) there exist
several inversions in $<v>$ inside regions of the same locking
characteristics, which means regions where $v$ is periodic with the same
period conmensurate with $T$, (see for example four of such inversions in $%
\Gamma $ between 1.01 and 1.015); b) there are regions of different locking
characteristics and the same mean velocity (see for example the region with $%
\Gamma $ between 1.015 and 1.035 and the region near $\Gamma $=1.045); c)
the values of $\Gamma$ where the inversions in $<v>$ take place are strongly
dependent on initial conditions.

To understand these characteristics, the trajectories are shown in figures 2
to 4. In figures 2a and 2b we show the case of two trajectories
corresponding to different locking zones ($\Gamma $=1.015 and $\Gamma $%
=1.035). Both trajectories (2a and 2b) show the same net transport given by
a straight line with slope $<v>=v_\omega $ but the oscillations $\tilde {x}$
over this straight line are different as can be seen in figures 2c and 2d
where the phase spaces ($\tilde {x}/T_x ,v/v_\omega$) are shown.

In figure 3 the case of a mean velocity reversal inside the period-1 zone is
shown; the complete trajectories for $\Gamma $=1.01 and $\Gamma $=1.011 are
drawn in 3a and 3b respectively. In 3c and 3d we show the transitory with
greater detail.

Finally in figure 4 the case $\Gamma $=1.527 corresponding to chaotic motion
is analyzed. In figure 4a we have plotted the trajectory $x(t)$ showing
clearly that we are in the presence of a net-transport phenomenon and in 4b
we show the phase space with the chaotic oscillations superimposed.

The above results clearly show that with method I the bifurcation diagram
gives the behavior of the oscillations superimposed on the mean motion and
not the net-transport movement. The phase locking analysis is a useful tool
for the study of the synchronization between auto-oscillatory systems with
an external periodic driving force\cite{13}. In that case the variables $x$
and $v$ are both periodic (i.e. S$^1$) and the net transport is
automatically discarded.

Figs. 5 and 6 show the normalized mean velocity and the bifurcation diagram
obtained with method II, which means taking last position and velocity of
the trajectory of the previous $\Gamma$ as the initial condition for the new 
$\Gamma$. The step $\Delta \Gamma$ is $0.001$ and the particle evolves
during $499 T$ for each $\Gamma$. Fig. 5 was obtained beginning with $\Gamma$%
=0.89665 as the initial $\Gamma$. The normalized mean velocity remains
unchanged equal to 1 until $\Gamma$ approximately 1.05. Fig. 6 was obtained
beginning with $\Gamma$=0.89666 as the initial $\Gamma$. The normalized mean
velocity remains unchanged at -1 until $\Gamma$ is approximately 1.08. The
initial $\Gamma$ was chosen so that there is a current reversal between them
as can be seen in Fig. 1 c). In Figs. 5a) and 6a) there is no current
reversal in the range $[0.9,1.05]$, while there are many current reversals
in the same zone of Fig. 1 a). On the other hand, the bifurcation diagrams
of Figs. 5b) and 6b) are continuous but different from each other,
furthermore none of them shows the abrupt changes appearing in the same zone
of Fig. 1b).

These results may be explained if there are two attractors corresponding to
mean velocities $+v_\omega$ and $-v_\omega$. In each of the simulations
obtained by method II the particle remains locked into one of the
attractors. The current reversals of Fig. 1 are due to the fact that the
border between the basins of attraction changes with $\Gamma$. For $\Gamma$
in some region around $\Gamma$=1 the mean velocity may be positive or
negative depending on which basin of attraction the initial condition
belongs to. This sensitivity to the initial condition may be used to control
current reversals by means of a small perturbation in $\Gamma$ when the
particle is near the border of a basin of attraction. An experiment may be
envisioned where a digital generator is used as the external force, in that
case the step in $\Gamma $ and the precise time when the change is applied
may be tuned to control the trajectory of each particle. As an example in
Fig. 7 we show the case of a particle starting at $x_0$=0 and $v_0$=1 with $%
\Gamma =0.89665$. This particle reaches a final value of $<v>/v_\omega $=1
as can be seen in Fig. 1. The bold curve in Fig. 7 shows the effect of
changing $\Gamma$ to 0.89675 at time $t=498.7 T$ , when the particle is at $%
x=3.1097+2n\pi $ with an instantaneous velocity $v =0.911662$. Due to this
change, movement of the particle is reversed. On the contrary, if the same
change in $\Gamma $ is produced at $t=499 T$, when the particle is at $%
x=1.0597+2n\pi $ with an instantaneous velocity $v=-1.82108$, the inversion
does not occur as is shown with the thin curve.

These above results were obtained by studying the trajectory of only one
particle, but ratchet transport is essentially stochastic. This was already
pointed out by Feynman \cite{Feynman}, when he used the ratchet to introduce
the Second Law of Thermodynamics. In deterministic inertial ratchets
deterministically induced chaos mimics the role of noise \cite{13a} and this
calls for the use of a time dependent probability measure, as has been
previously done in \cite{9,quenched,quenched2}.

We use a collection of particles with different initial conditions to study
the transport phenomena. We work with an ensemble of $N$ particles having
identical initial velocities $v_0$ but initial positions equally distributed
in the range $[x_{\min }, x_{\max }]$. The initial probability density is
given by:

\[
\rho (x,v,0)=\delta (v-v_{0})\left[ H(x-x_{\min })-H(x-x_{\max })\right] , 
\]
where $H(x)$ is the step function.

The normalized mean-velocity of the ensemble is:

\[
<<V>>=\frac {1}{N }\sum_{i=1}^N<v>_i. 
\]

As for the one particle case we first show the results of simulations with
method I, which in the packet of particles means returning to the initial
condition $\rho (x,v,0)$ when $\Gamma$ changes.

The results shown in figure 8 were obtained with $N$=200 and $[x_{\min }$,$%
x_{\max }]=[5.08,11.35]$ and $v_{0}=0.01$ corresponding to a well spread-out
packet of particles with initial positions between two maxima of the
potential. The corresponding bifurcation diagram is also shown. There is
only one current reversal at $\Gamma \simeq 1.05$, where an order-disorder
transition takes place. About three quarters of the particles in the
ensemble have mean velocity $<v>_i=-v_w$ and the remaining quarter have $%
<v>_i=v_w$ giving a normalized ensemble mean velocity $<<V>>/v_\omega \simeq
-0.5$. These results seem to agree with Mateos' conjecture\cite{10} that
current inversion is associated with order-disorder transition in the
bifurcation diagram.

In order to obtain the behavior corresponding to a narrow initial packet, we
work with the initial condition $[x_{\min }, x_{\max }]= [ 5.08, 5.09]$ and $%
v_{0}=0.01$. In this ensemble particles are initially located near the
maximum of the potential and having a small initial velocity. Fig. 9 shows a
current inversion around $\Gamma =1.015.$ The corresponding region at the
bifurcation diagram has no order-disorder transition, contradicting Mateos'
conjecture. However, the same current reversal, associated with an
order-disorder transition, which was obtained for the case of a wide packet
for $\Gamma \simeq 1.05$, does takes place.

Initial sets of particles with positions near the minimum of the potential
and sets of particles with identical initial positions and different initial
velocities equally distributed were also studied with qualitatively similar
results.

It is possible to use method II to control the normalized mean velocity of
the packet. For example to obtain a normalized mean velocity $<<V>>/v_\omega
= -0.5$ which is the minimum normalized mean velocity of Fig. 9 a), with the
same narrow packet used above we use method II starting with $%
\Gamma_{min}=1.013$ and increasing it up to $\Gamma=1.06$ in 100 equal
steps. The time period for each step in $\Gamma$ was conmensurate with $T$.
The normalized mean velocity as a function of $\Gamma$ is shown in Fig. 10
a). The corresponding bifurcation diagram is shown in Fig. 10b). As $\Gamma$
changes each particle remains locked to its mean velocity $<v>_i$
corresponding to $\Gamma_{min}$ and the packet also remains locked to its
initial mean velocity. This behavior persists until $\Gamma = 1.0495$ where
the mean velocity drops to $<<V>>/v_\omega = -1.$ The unlocking effect
corresponds to a chaos to order transition. If the simulation starts with
any $\Gamma_{min}$ producing a positive velocity locking for the packet the
unlocking effect produces a current reversal as Mateos found.

In summary, we studied the variation of the mean velocity of one particle
and of a packet of particles with the amplitude of the external force $%
\Gamma $ in two ways. The first one consists of returning to the initial
conditions whenever $\Gamma$ is changed. In this way many current reversals
appear which are not associated with the bifurcations in the bifurcation
diagram. The second way is to take as initial condition the last position
and velocity of the previous trajectory for a given $\Gamma$. When this
method is used, the mean velocity remains locked in the periodic zones of
the bifurcation diagram. The differences between the results obtained with
both methods may be explained by the presence of at least two attractors
associated with positive and negative mean velocities. The two methods are
complementary as possible mechanisms for the control of current reversal,
because method I reveals the existence of different attractors and gives
information about their basins of attraction, while method II, although
history dependent, shows the locking process.

We conclude that control of current reversals in deterministic inertial
ratchets is possible as a consequence of a locking process associated with
different mean velocity attractors. An unlocking effect is produced when a
chaos to order transition limits the control range.

This work was supported by grants from the Office of Naval Research and from
the Universidad Nacional de Mar del Plata. We acknowledge Mihail N. Popescu
for very useful discussions on simulation methods and control. C.M.A.
acknowledges Alvaro L. Salas Brito for useful discussions.

%\section{Bibliography (atenci\'{o}n hay que elegir en los 8 primeros)}

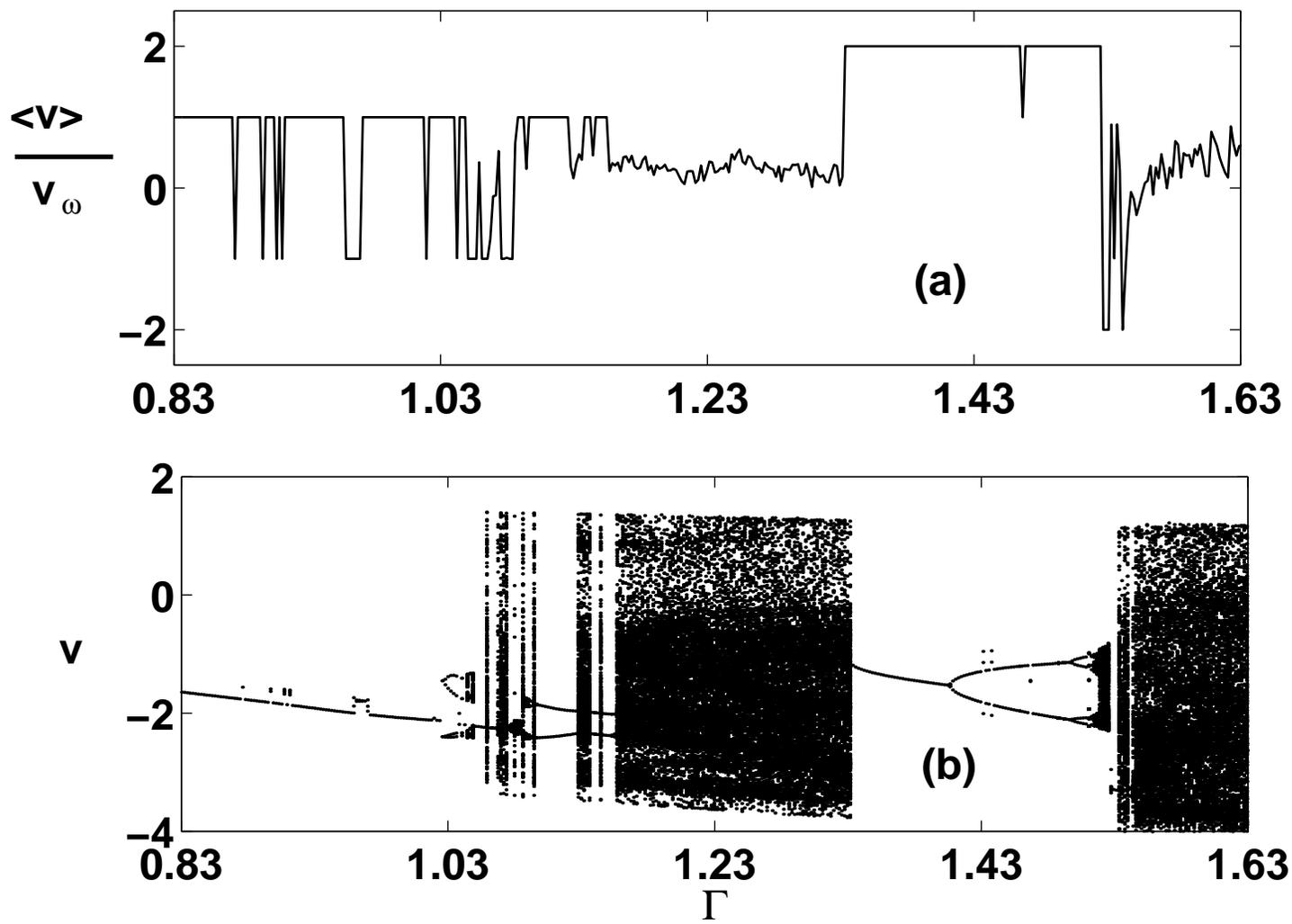
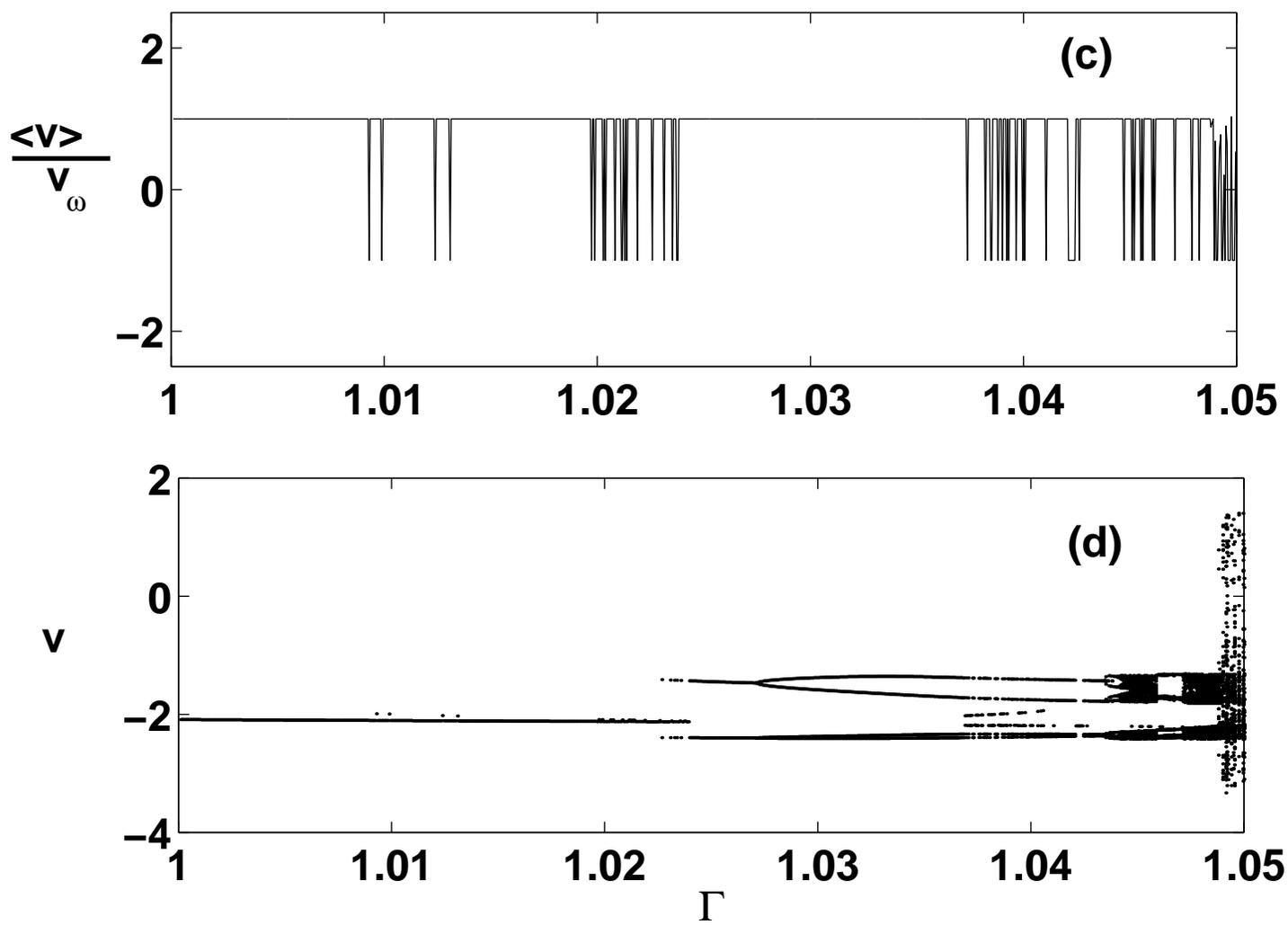
\begin{figure}[tbp]
\caption{a) Normalized mean velocity of a particle with initial condition $%
x_0=0 , v_0=1$ as a function of $\Gamma$. Initial condition is reset
whenever $\Gamma$ changes (Method I). b) Velocity bifurcation diagram for a
particle with initial condition $x_0=0 , v_0=1$ as a function of $\Gamma$.
Initial condition is reset whenever $\Gamma$ changes (Method I). c)
Enlargement of Fig. 1a) for $\Gamma$ between $[1,1.05]$ d) Enlargement of
Fig. 1b) for $\Gamma$ between $[1,1.05]$ }
\end{figure}

\begin{figure}[tbp]
\caption{a) Trajectory of a particle with initial condition $x_0=0 , v_0=1$
for $\Gamma = 1.015$ b) Trajectory of a particle with initial condition $%
x_0=0 , v_0=1$ for $\Gamma = 1.035$ c) Phase Space corresponding to
oscillations superimposed on the net motion of a particle with initial
condition $x_0=0 , v_0=1$ for $\Gamma = 1.015$ d) Phase Space corresponding
to oscillations superimposed on the net motion of a particle with initial
condition $x_0=0 , v_0=1$ for $\Gamma = 1.035$}
\end{figure}

\begin{figure}[tbp]
\caption{a) Trajectory of a particle with initial condition $x_0=0 , v_0=1$
for $\Gamma = 1.010$ b) Trajectory of a particle with initial condition $%
x_0=0 , v_0=1$ for $\Gamma = 1.011$ c) Enlargement of the transitory
corresponding to the trajectory of a particle with initial condition $x_0=0
, v_0=1$ for $\Gamma = 1.010$ d) Enlargement of the transitory corresponding
to the trajectory of a particle with initial condition $x_0=0 , v_0=1$ for $%
\Gamma = 1.011$}
\end{figure}

\begin{figure}[tbp]
\caption{a) Trajectory of a particle with initial condition $x_0=0 , v_0=1$
for $\Gamma = 1.527$ b) Phase Space corresponding to oscillations
superimposed to the net motion of a particle with initial condition $x_0=0 ,
v_0=1$ for $\Gamma = 1.527$}
\end{figure}

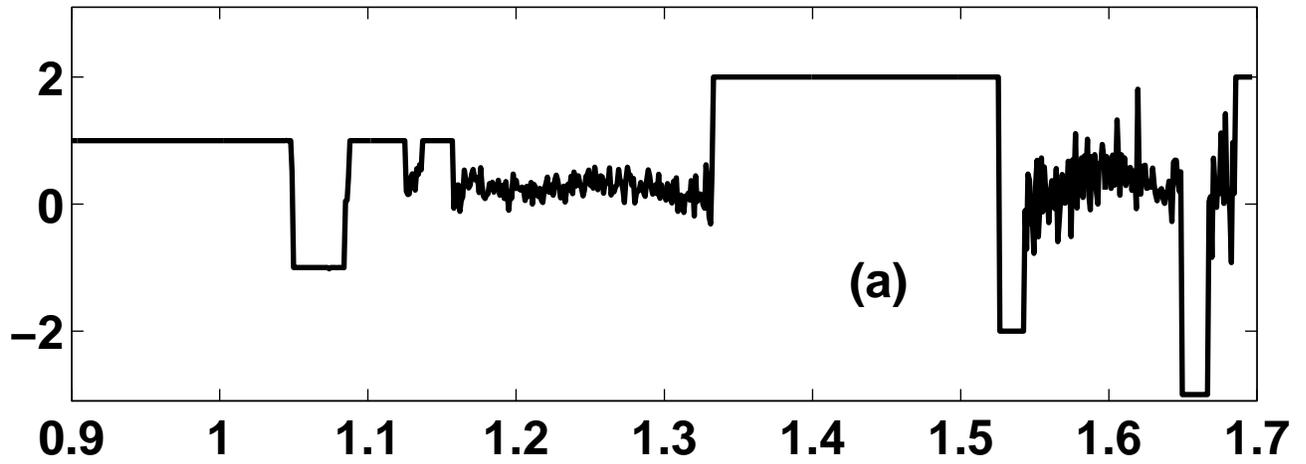
\begin{figure}[tbp]
\caption{a) Normalized mean velocity of a particle with initial condition $%
x_0=0 , v_0=1$ as a function of $\Gamma$ starting at $\Gamma = 0.89665$. The
initial condition for a given $\Gamma$ is the last position and velocity of
the trajectory of the previous $\Gamma$ (Method II). b) Velocity bifurcation
diagram for a particle with initial condition $x_0=0 , v_0=1$ as a function
of $\Gamma$ starting at $\Gamma = 0.89665$. The initial condition for a
given $\Gamma$ is the last position and velocity of the trajectory of the
previous $\Gamma$ (Method II).}
\end{figure}

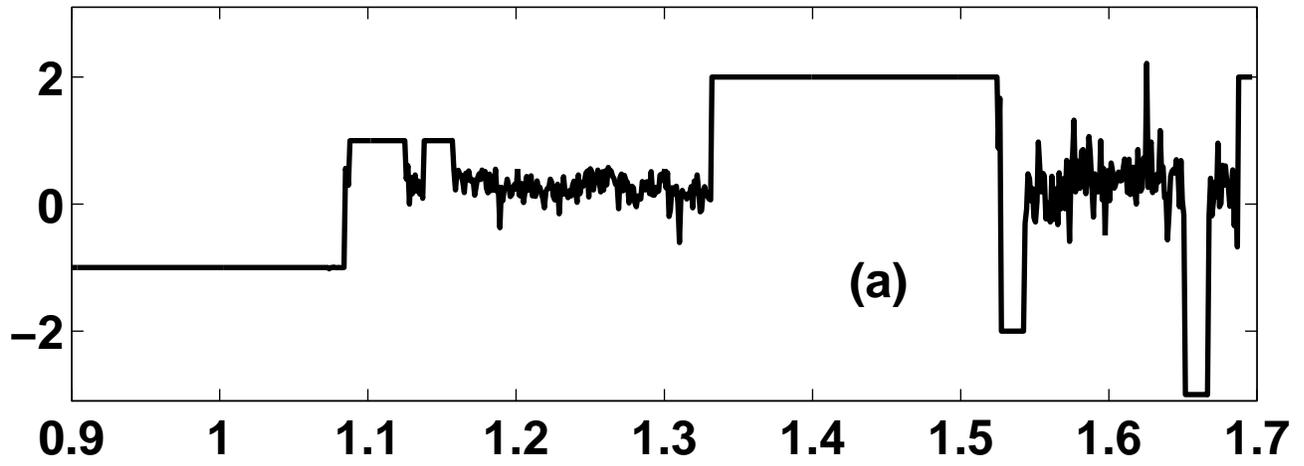
\begin{figure}[tbp]
\caption{a) Normalized mean velocity of a particle with initial condition $%
x_0=0 , v_0=1$ as a function of $\Gamma$ starting at $\Gamma = 0.89666$. The
initial condition for a given $\Gamma$ is the last position and velocity of
the trajectory of the previous $\Gamma$ (Method II). b) Velocity bifurcation
diagram for a particle with initial condition $x_0=0 , v_0=1$ as a function
of $\Gamma$ starting at $\Gamma = 0.89666$. The initial condition for a
given $\Gamma$ is the last position and velocity of the trajectory of the
previous $\Gamma$ (Method II).}
\end{figure}

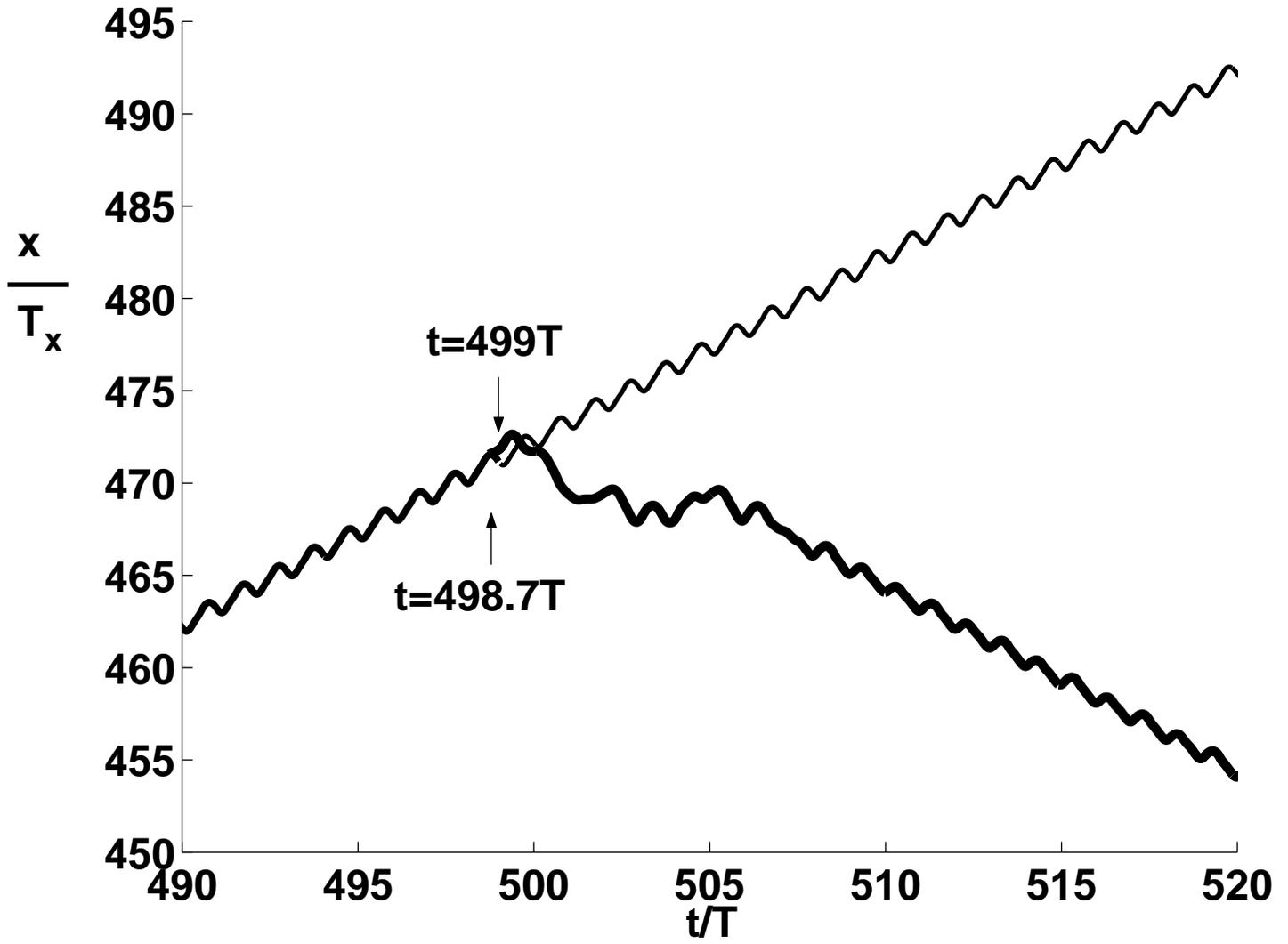
\begin{figure}[tbp]
\caption{Example of control of current direction by selecting the time when $%
\Gamma$ is changed from $\Gamma = 0.89665$ to $\Gamma = 0.89666$. The bold
curve corresponds to a changing time $t= 498.7 T$. The thin curve
corresponds to a changing time $t = 499 T$. }
\end{figure}

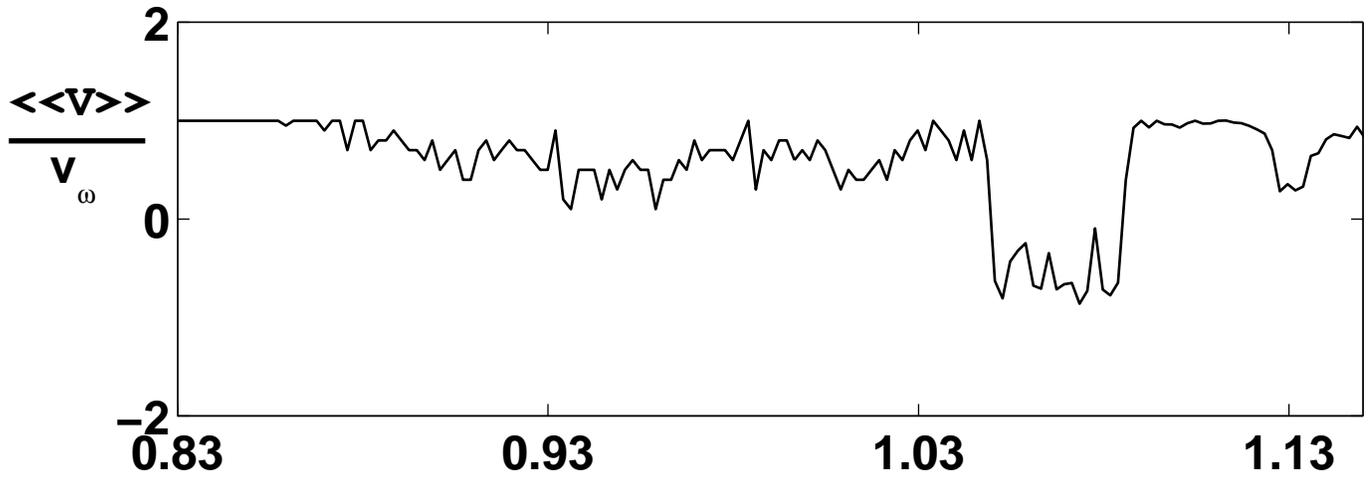
\begin{figure}[tbp]
\caption{a) Mean velocity of a set of particles with initial positions
between two maxima of the potential $[5.08,11.35]$ and initial velocity $v_0
= 0.01$ as a function of $\Gamma$. Initial conditions are reset whenever $%
\Gamma$ changes (Method I). b) Velocity bifurcation diagram for a particle
of the set of particles with initial positions between two maxima of the
potential $[5.08,11.35]$ and initial velocity $v_0 = 0.01$ as a function of $%
\Gamma$. Initial conditions are reset whenever $\Gamma$ changes (Method I).}
\end{figure}

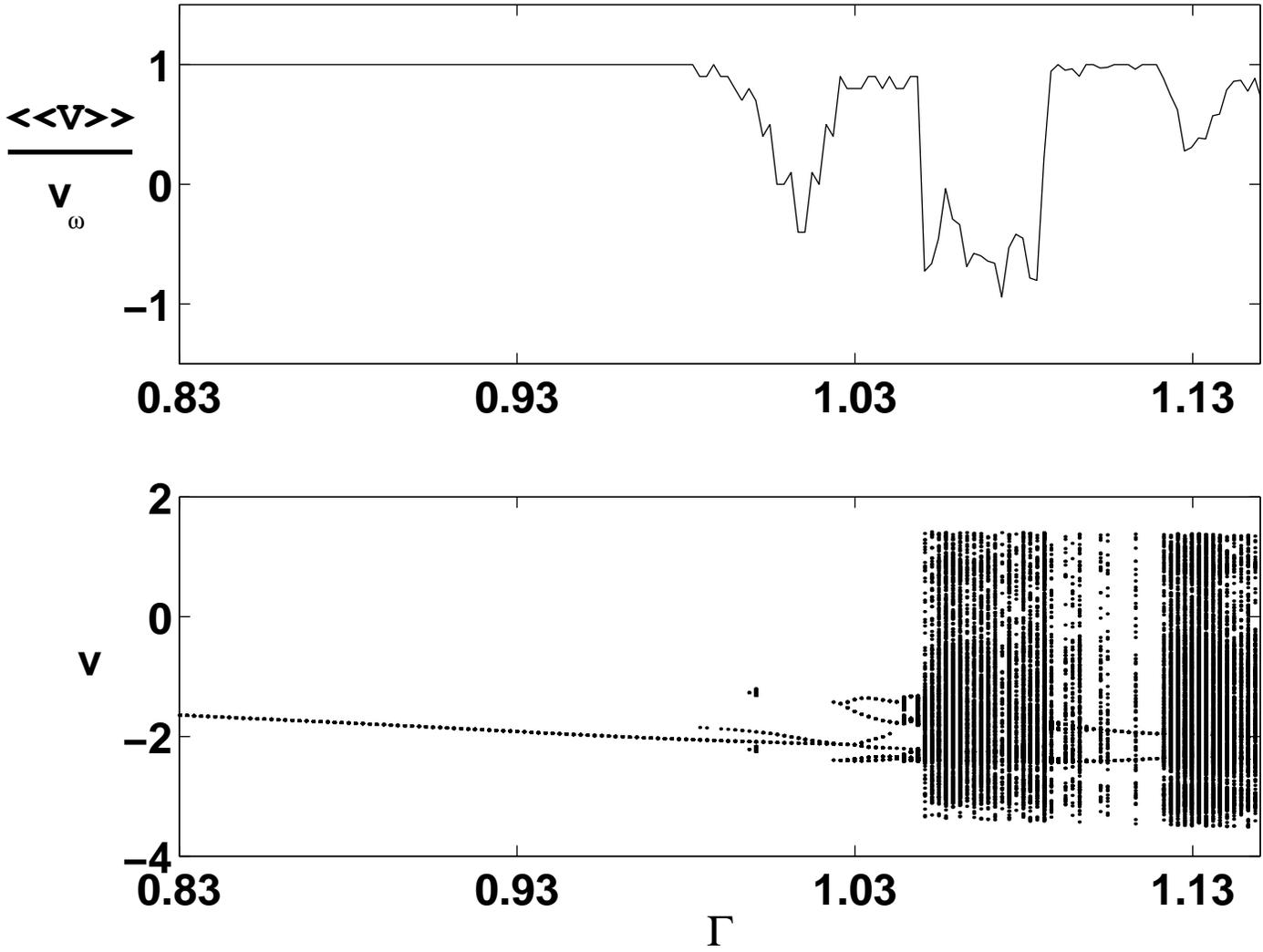
\begin{figure}[tbp]
\caption{a) Mean velocity of a narrow set of particles centered at the
maximum of the potential and initial velocity $v_0 = 0.01$ as a function of $%
\Gamma$. Initial conditions are reset whenever $\Gamma$ changes (Method I).
b) Velocity bifurcation diagram for a particle of the set of particles
centered at the maximum of the potential and initial velocity $v_0 = 0.01$
as a function of $\Gamma$. Initial conditions are reset whenever $\Gamma$
changes (Method I).}
\end{figure}

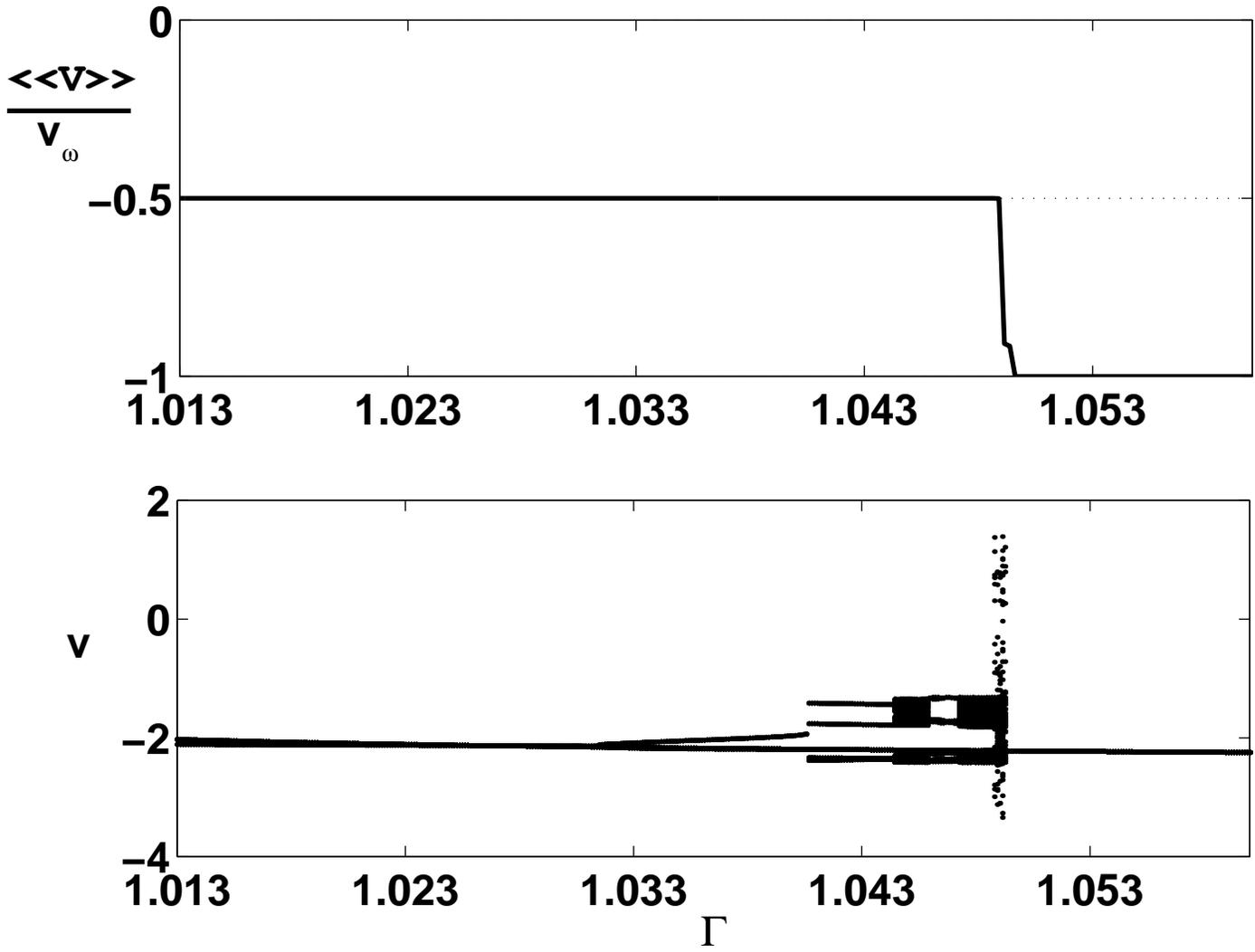
\begin{figure}[tbp]
\caption{a) Example of control of mean velocity for a narrow distribution of
particles centered at the maximum of the potential and initial velocity $v_0
= 0.01$ as a function of $\Gamma$ starting at $\Gamma=1.013$. The initial
conditions for a given $\Gamma$ are the last positions and velocities of the
trajectories of the previous $\Gamma$ (Method II). b) Velocity bifurcation
diagram for a particle of the set of particles centered at the maximum of
the potential and initial velocity $v_0 = 0.01$ as a function of $\Gamma$
starting at $\Gamma=1.013$. The initial conditions for a given $\Gamma$ are
the last positions and velocities of the trajectories of the previous $%
\Gamma $ (Method II). }
\end{figure}

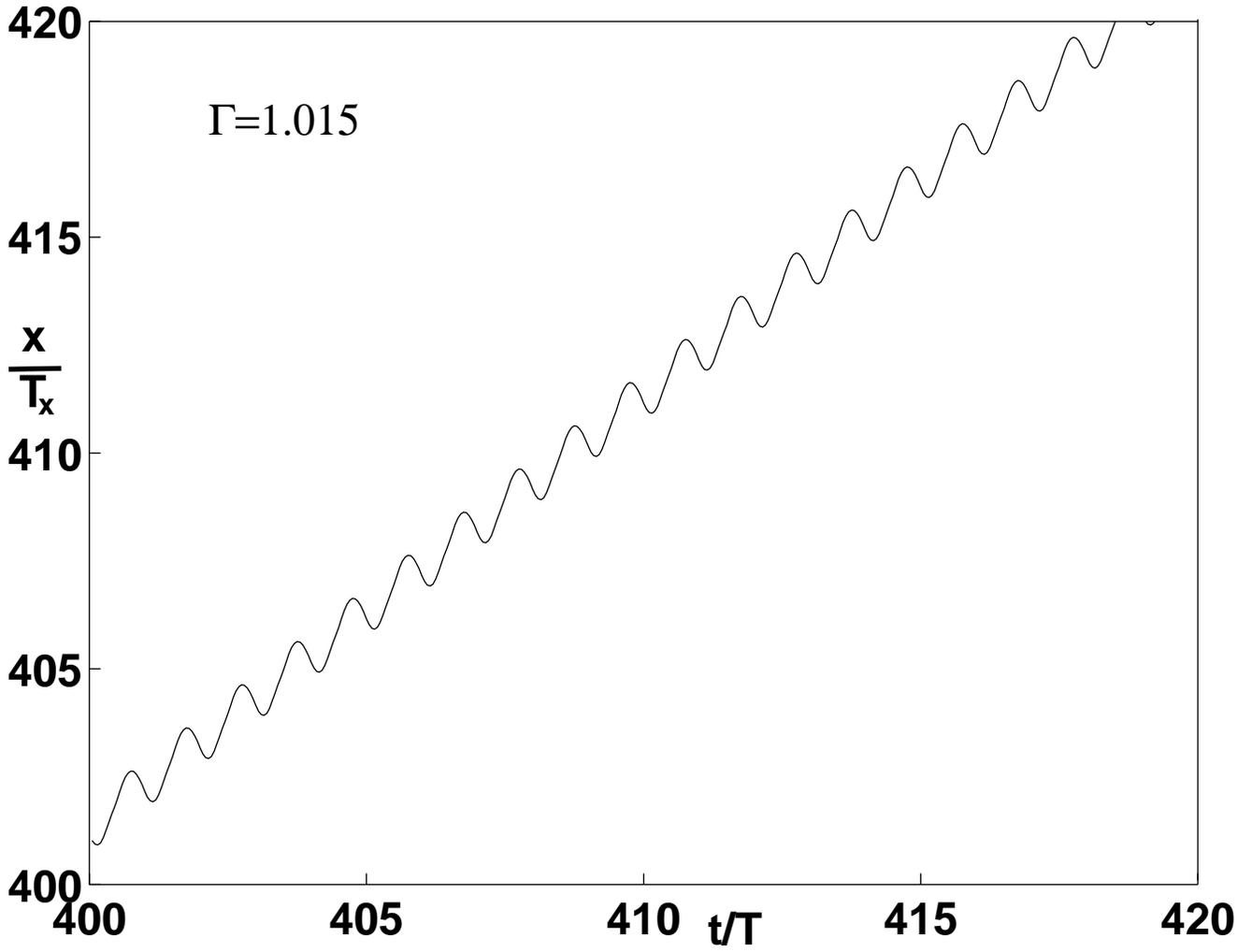
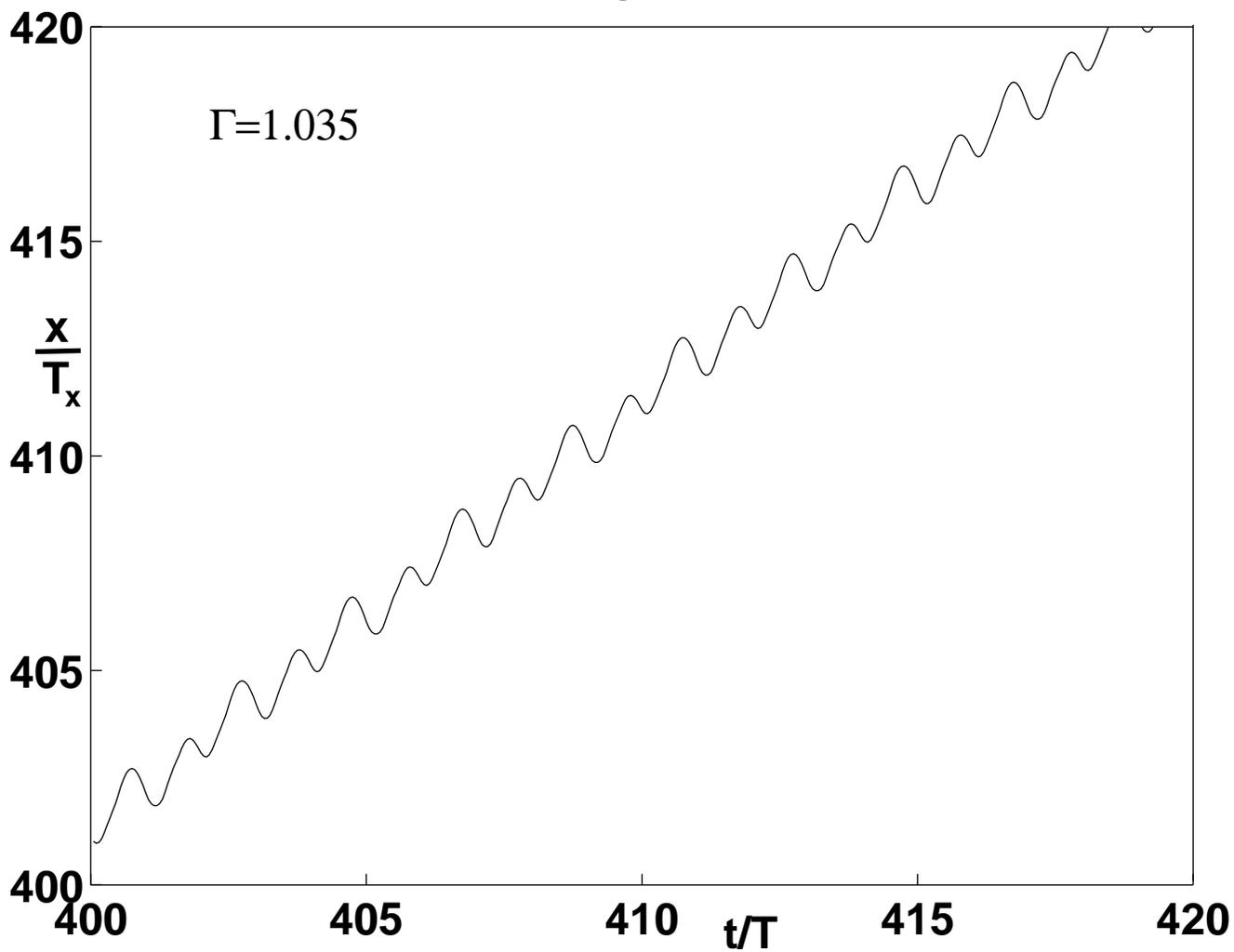
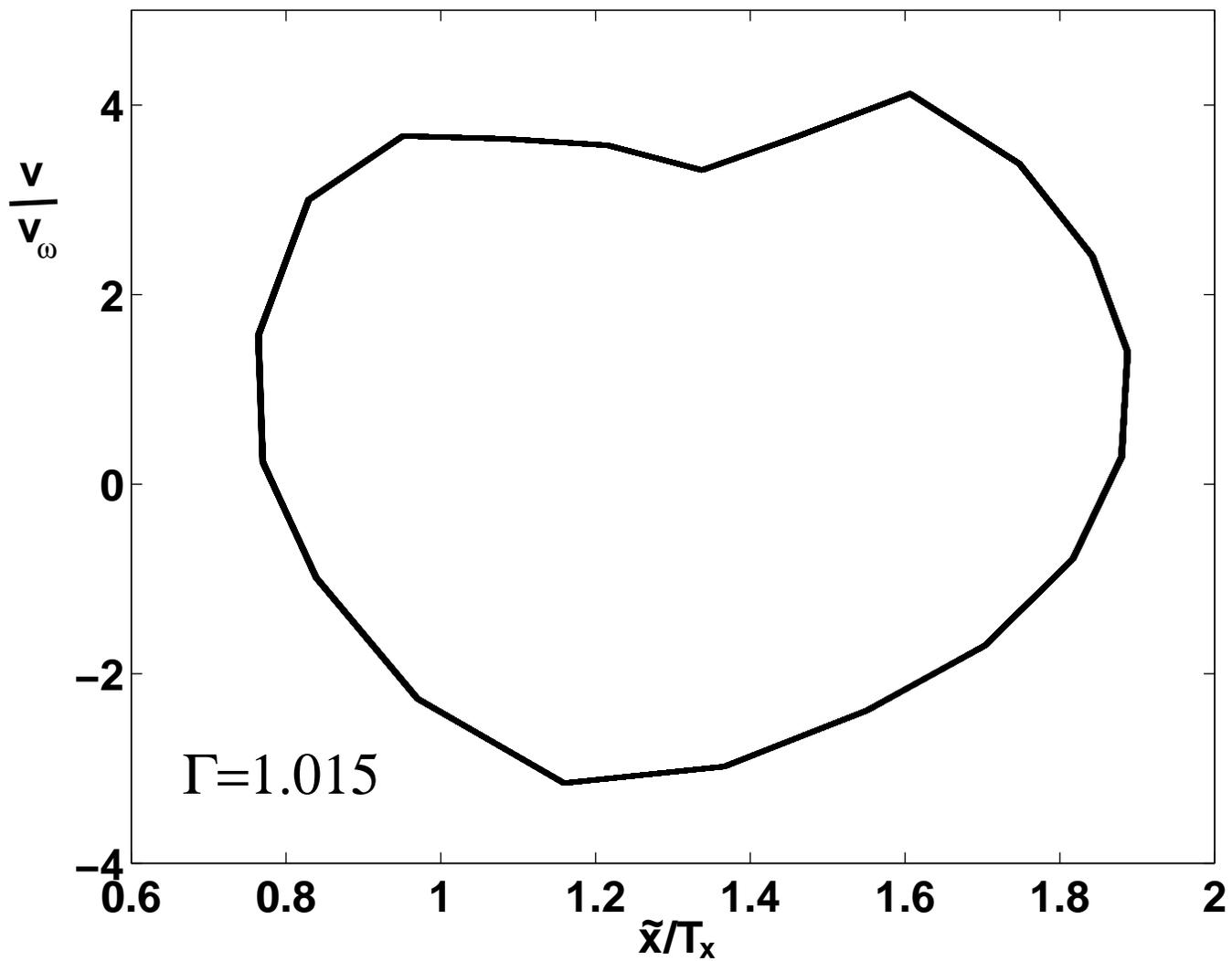
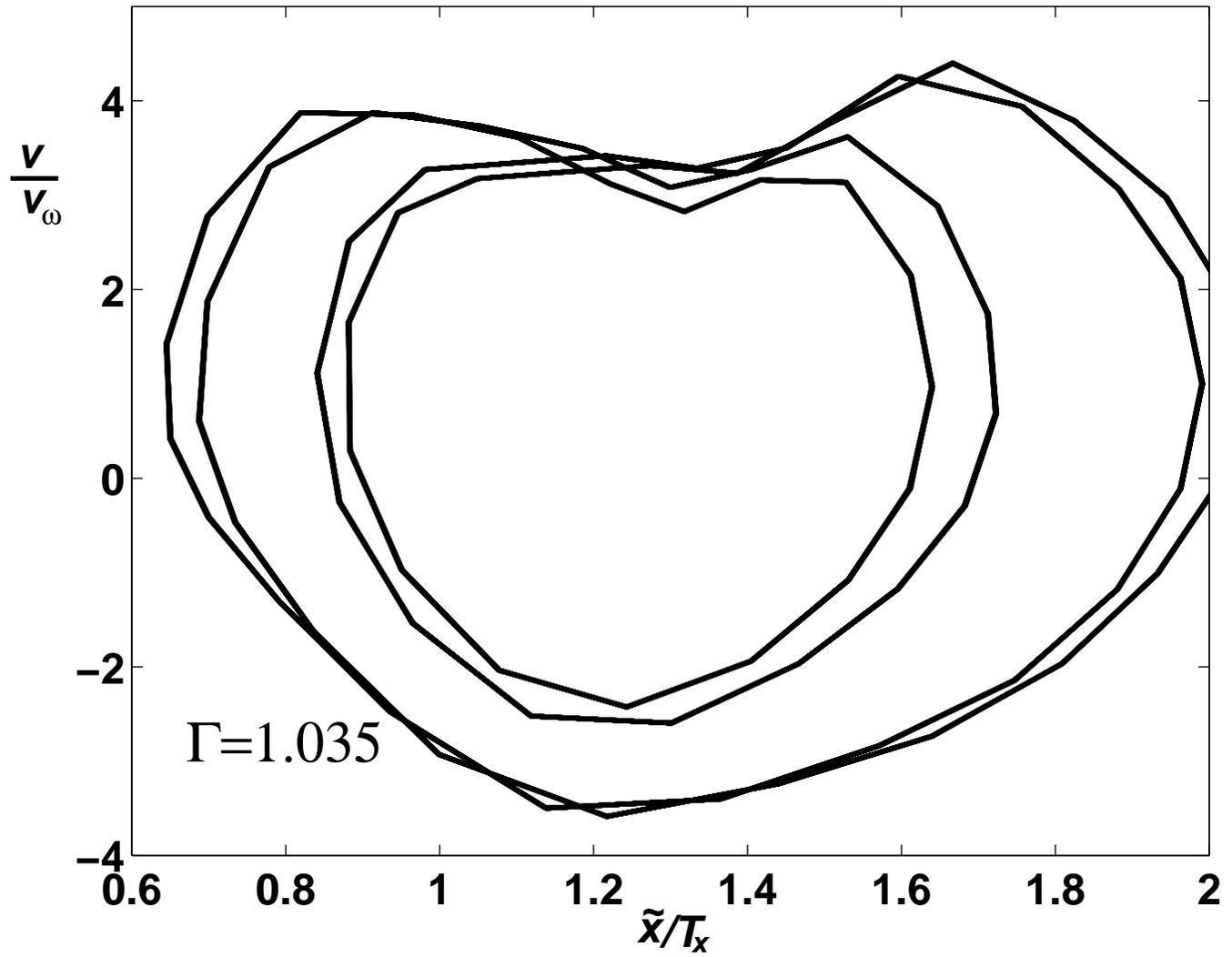
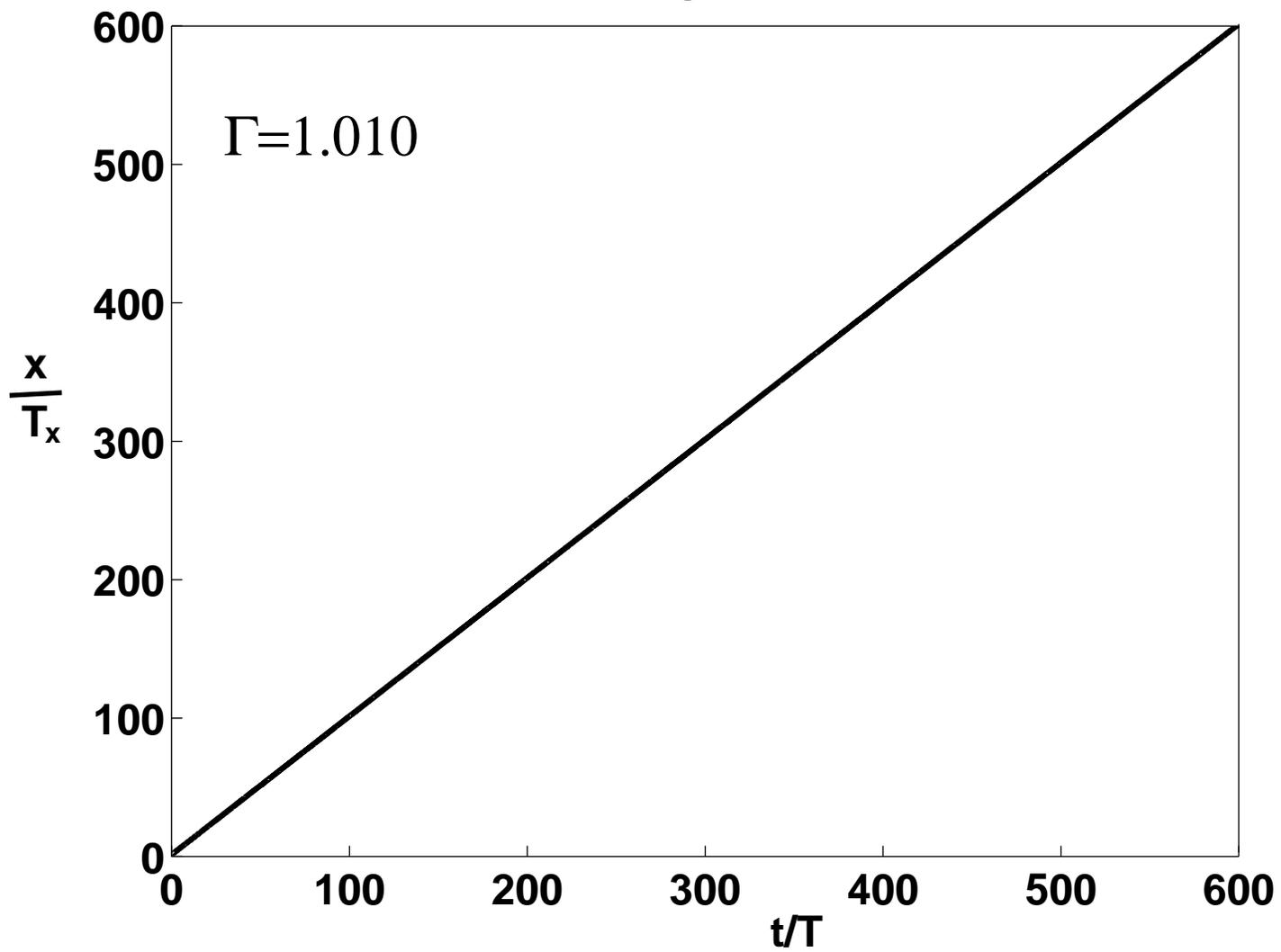
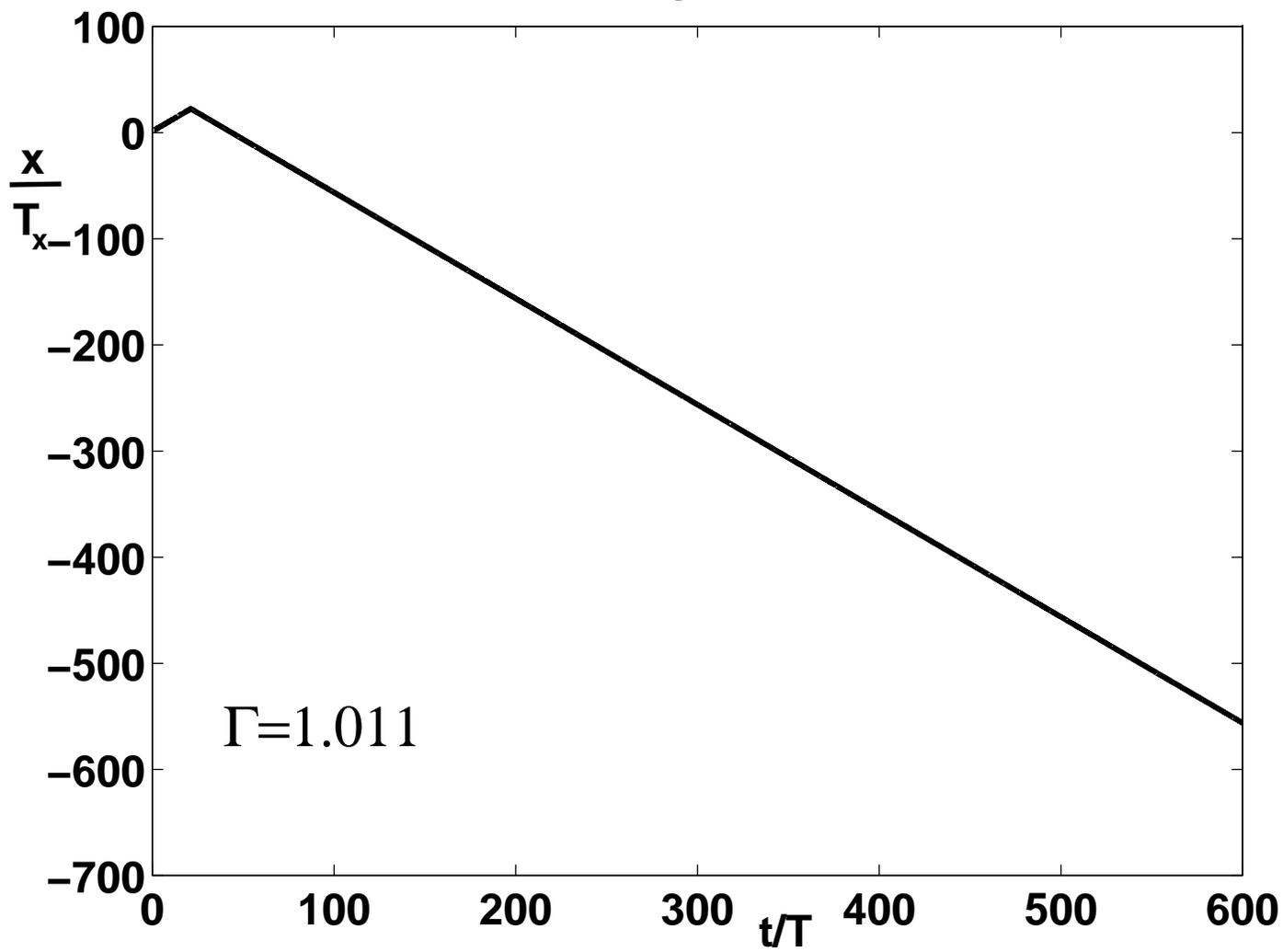
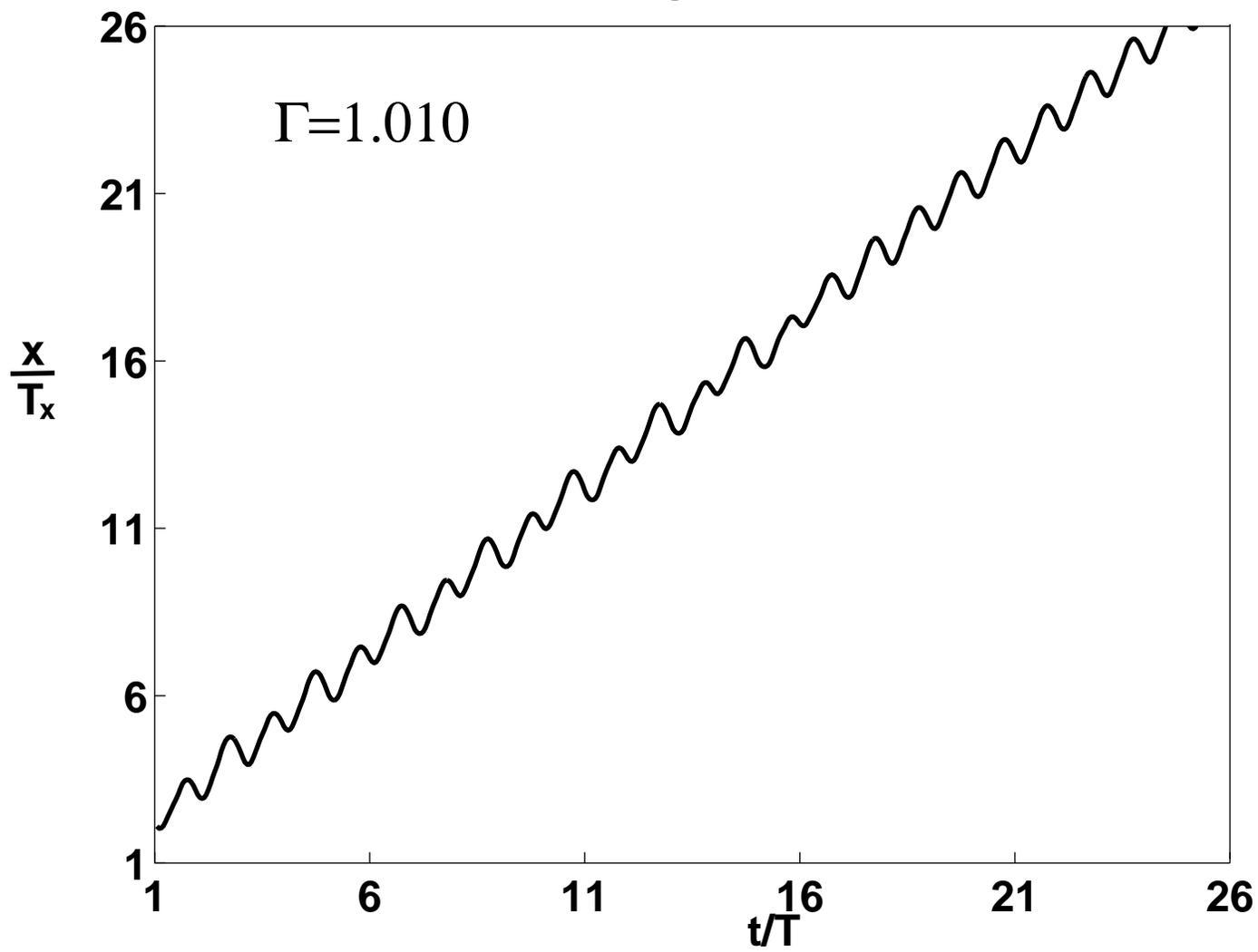
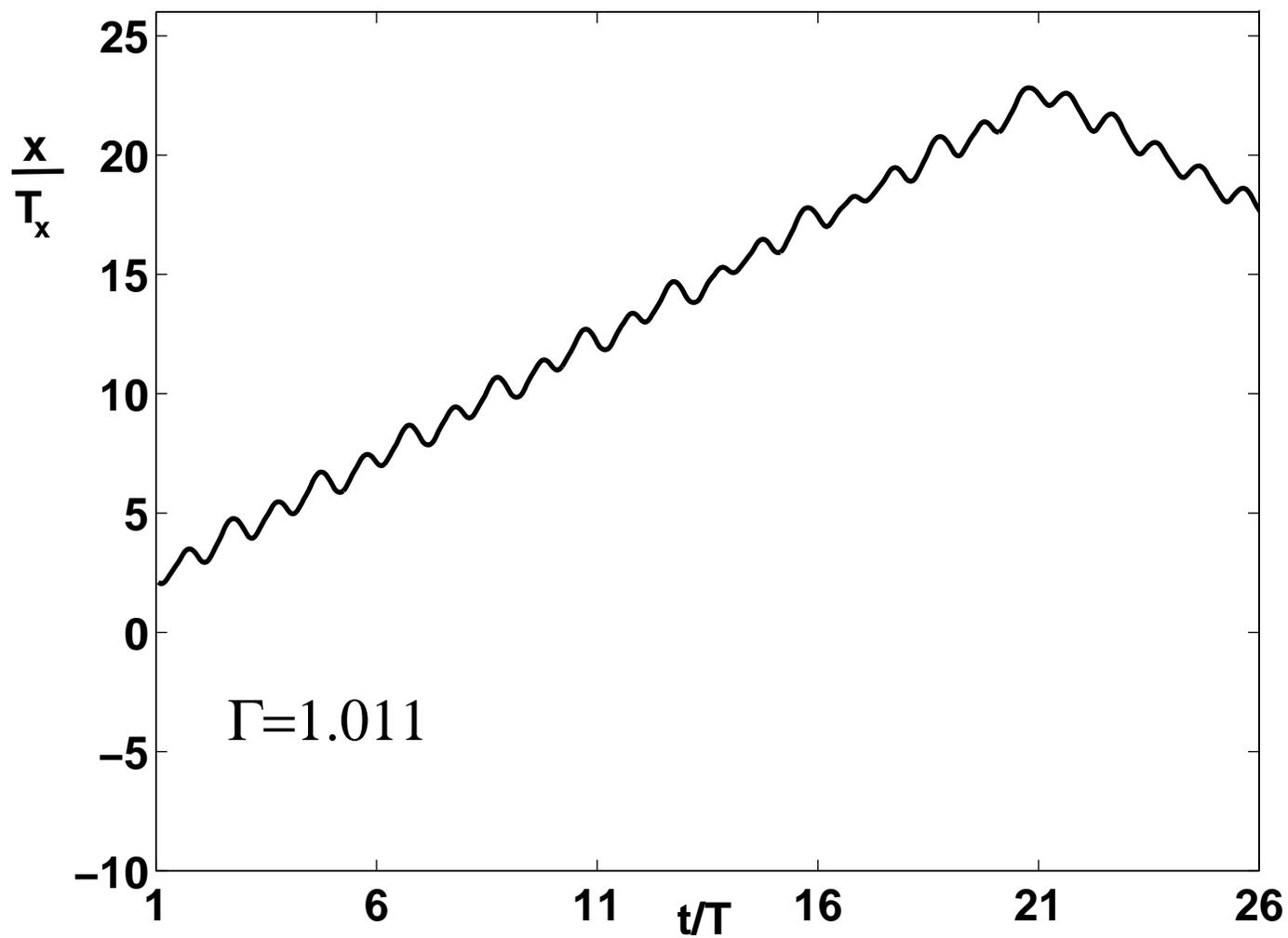
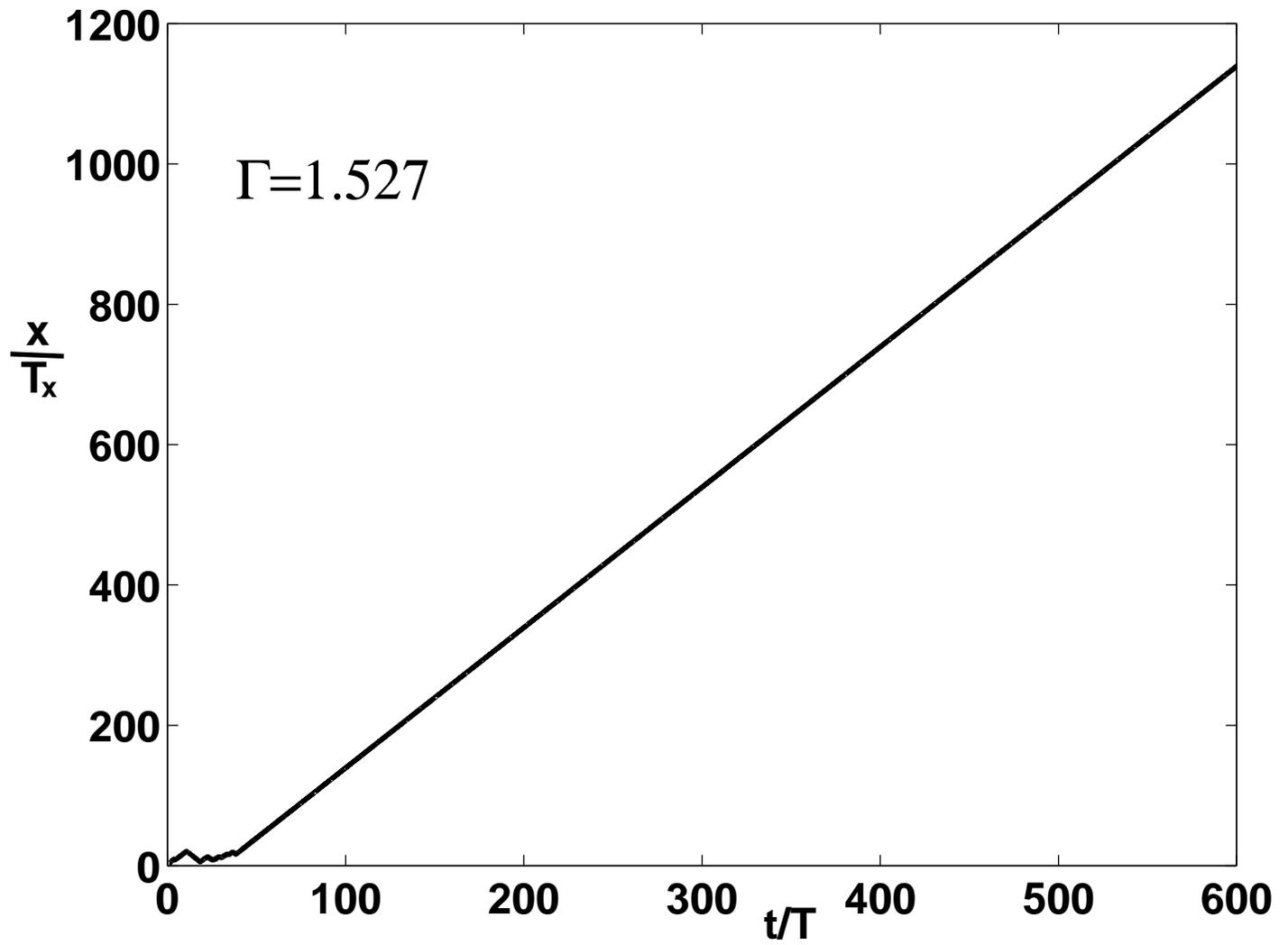
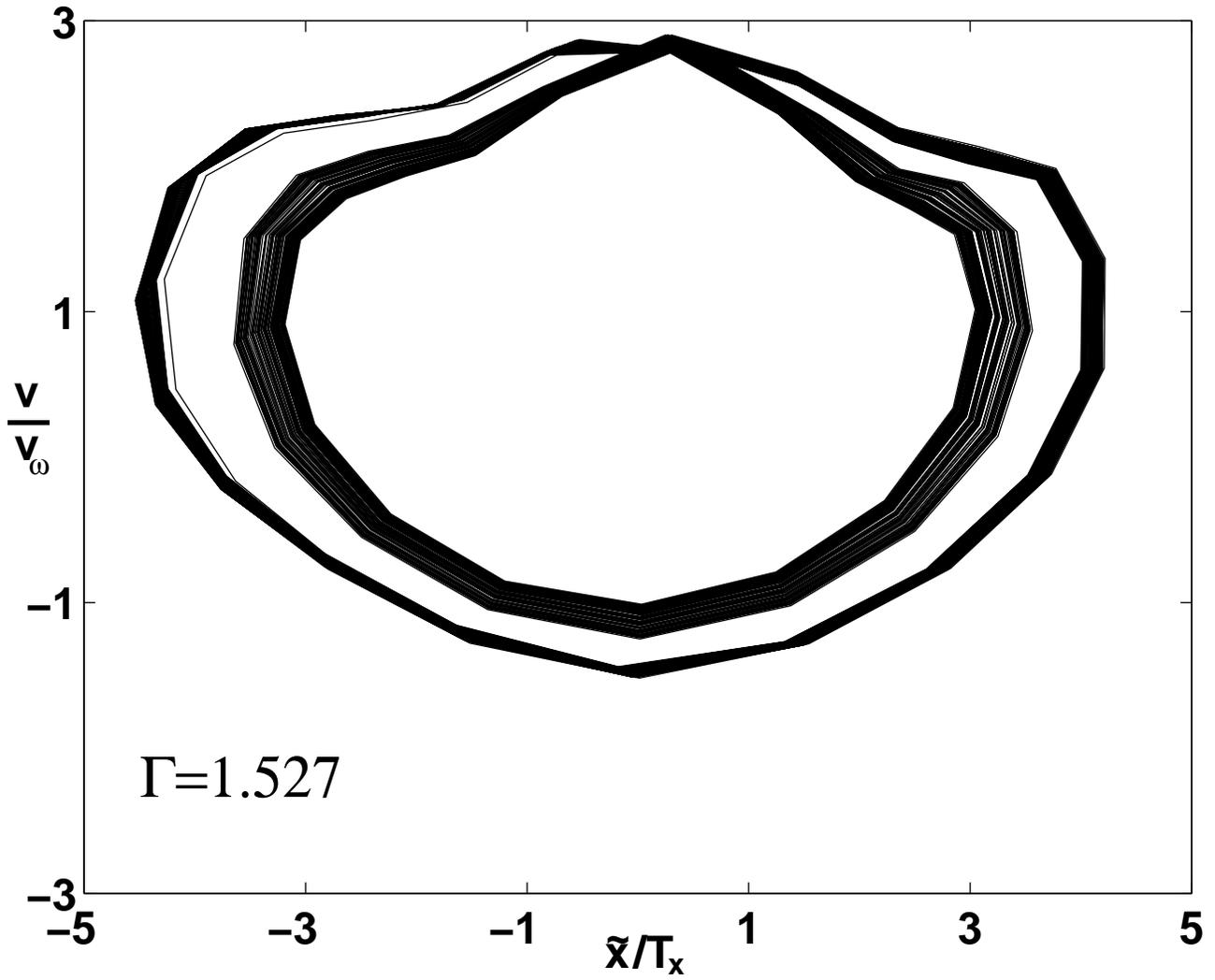
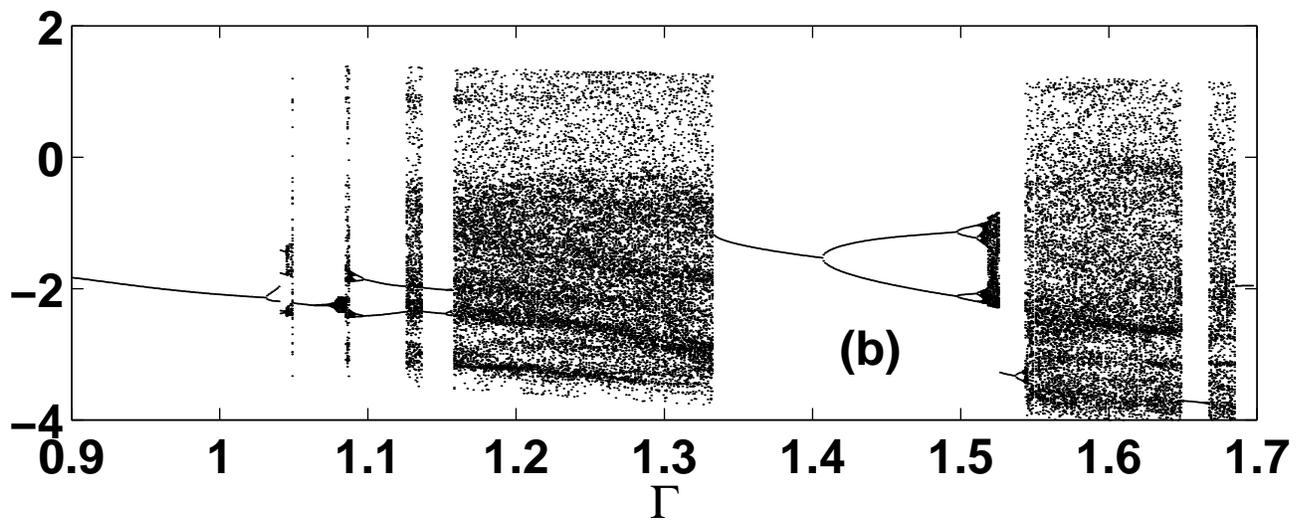
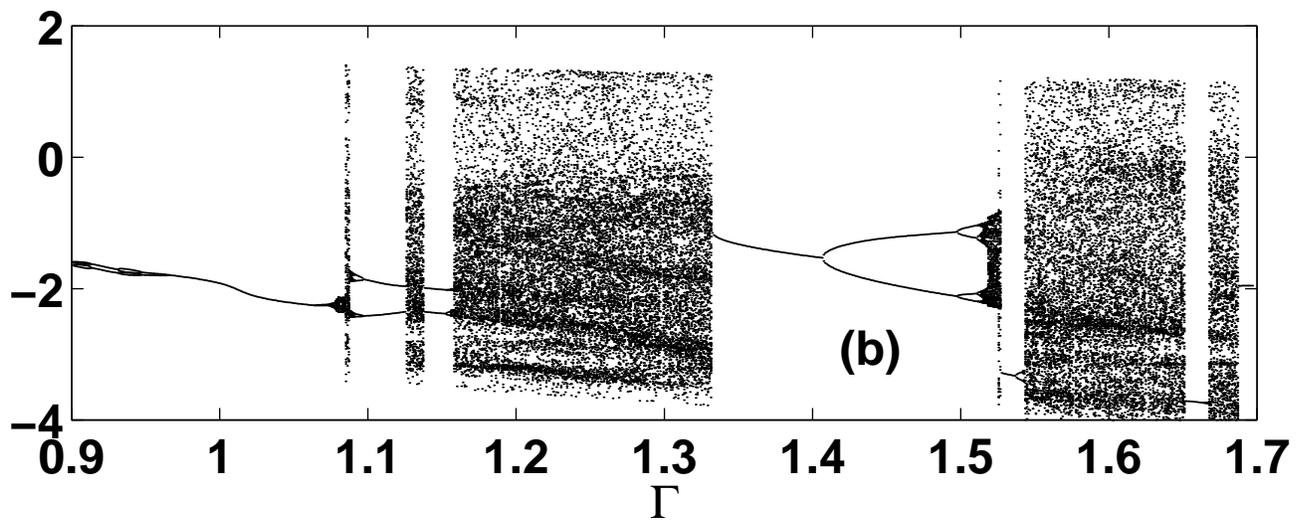
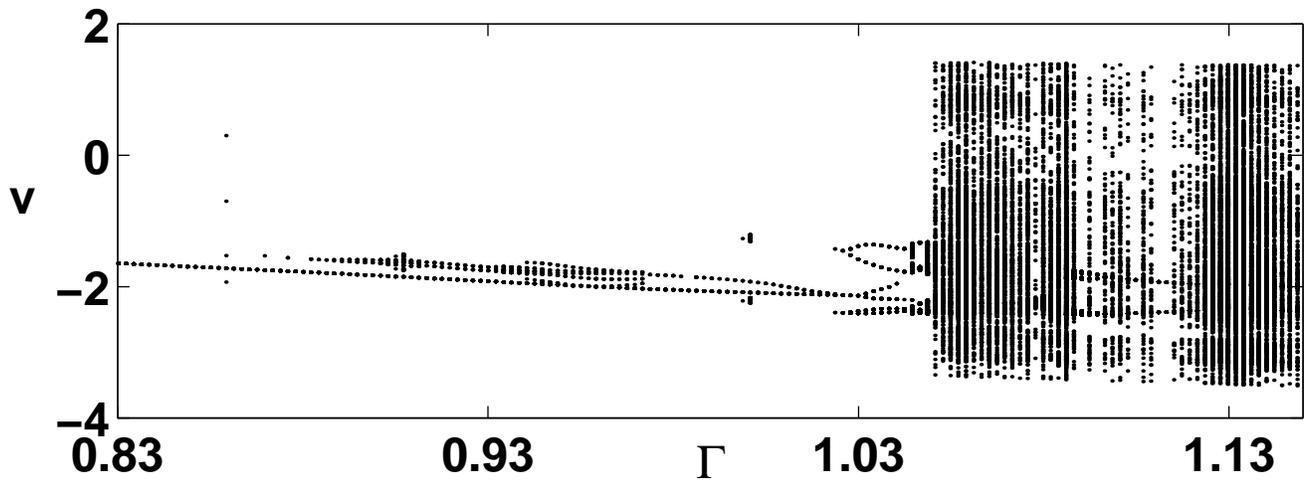
\end{document}